\documentclass[prd,twocolumn,superscriptaddress,nofootinbib,floatfix,preprintnumbers]{revtex4}
\pdfoutput=1

\usepackage{amssymb, bm, amsmath, graphicx}
\usepackage[colorlinks=true,linktocpage=true,linkcolor=blue,citecolor=blue]{hyperref}
\usepackage[usenames,dvipsnames]{color}
\usepackage{slashed}
\usepackage[utf8]{inputenc}
\usepackage[T1]{fontenc}
\usepackage{enumerate}
\usepackage{amsfonts}
\usepackage{mathtools}
\usepackage{latexsym}
\usepackage{hyperref}
\usepackage{epstopdf} 
\usepackage{bbm}
\usepackage{soul}
\usepackage[normalem]{ulem}
\usepackage{braket}
\usepackage{nccmath}
\usepackage{graphicx}
\usepackage{float}

\usepackage[e]{esvect}

\newcommand{\secn}[1]{Section~1}
\newcommand{\appn}[1]{Appendix~1}

\long\def\comment#1{ }

\def\and{\quad\text{and}\quad}

\def\0{{\boldsymbol 0}}
\def\1{{\boldsymbol 1}}
\def\p{{\boldsymbol p}}

\def\k{{\boldsymbol k}}
\def\x{{\boldsymbol x}}

\def\b{{\boldsymbol b}}
\def\0{{\boldsymbol 0}}

\def\bn{{\boldsymbol n}}

\renewcommand\a{\alpha}
\renewcommand\b{\beta}
\renewcommand\d{\delta}

\renewcommand\t{\tau}

\renewcommand\o{\omega}

\newcommand\m{\mu}
\newcommand\n{\nu}

\newcommand\D{\Delta}

\newcommand\pt{\partial}

\renewcommand{\part}{{\rm part}}

\newcommand{\be}{\begin{equation}}
\newcommand{\ee}{\end{equation}}
\newcommand{\bes}{\begin{subequations}}
\newcommand{\ees}{\end{subequations}}
\newcommand{\bea}{\begin{eqnarray}}
\newcommand{\eea}{\end{eqnarray}}

\newcommand{\nn}{\nonumber \\}

\begin{document}
\preprint{CERN-TH-2024-180}

\title{Giving wake to energy-energy correlators: \\hydrodynamic response on the celestial sphere}


\author{Jo\~{a}o Barata}
\email[Email: ]{joao.lourenco.henriques.barata@cern.ch}
\affiliation{CERN, Theoretical Physics Department, CH-1211 Geneva 23, Switzerland}

\author{Matvey V. Kuzmin}
\email[Email: ]{kuzmin.mv19@physics.msu.ru}
\affiliation {Physics Department, Lomonosov Moscow State University, 1-2 Leninskie Gory, Moscow 119991, Russia}

\author{José Guilherme Milhano}
\email[Email: ]{gmilhano@lip.pt}
\affiliation{LIP, Av. Prof. Gama Pinto, 2, P-1649-003 Lisboa, Portugal}
\affiliation{Departmento de Fisica, Instituto Superior Tecnico (IST), Universidade de Lisboa, Av. Rovisco Pais 1, P-1049-001 Lisboa, Portugal}

\author{Andrey V. Sadofyev}
\email[Email: ]{sadofyev@lip.pt}
\affiliation{LIP, Av. Prof. Gama Pinto, 2, P-1649-003 Lisboa, Portugal}

\begin{abstract}
The observation of the medium response generated by the propagation of high energy partons in the quark gluon plasma produced in heavy-ion collisions would provide a clear and unmistakable evidence for the hydrodynamic behavior of the bulk. Recently, it has been argued that the features of the medium's back-reaction to the jet could be cleanly imprinted in the correlations of asymptotic energy flows, in principle allowing to isolate this signal from other uncorrelated physical processes. Nonetheless, the current limited theoretical understanding of these jet observables in heavy-ion collisions 
constrains their applicability as probes of the medium (hydro)dynamics.
In this work, we provide an analytic picture for the medium back-reaction's effect on the energy flux and two point energy correlator.
We show that the 
medium response leads to the emergence of an universal classical scaling law, competing with the perturbative QCD contribution at large angles. Comparing the associated correlator to recent experimental measurements, we find that the observed large angle features can be qualitatively described by a purely hydrodynamically driven response and its interplay with the hard jet component.

\end{abstract}

\maketitle

\begin{figure}[t!]
    \centering
    \includegraphics[width=0.33\textwidth]{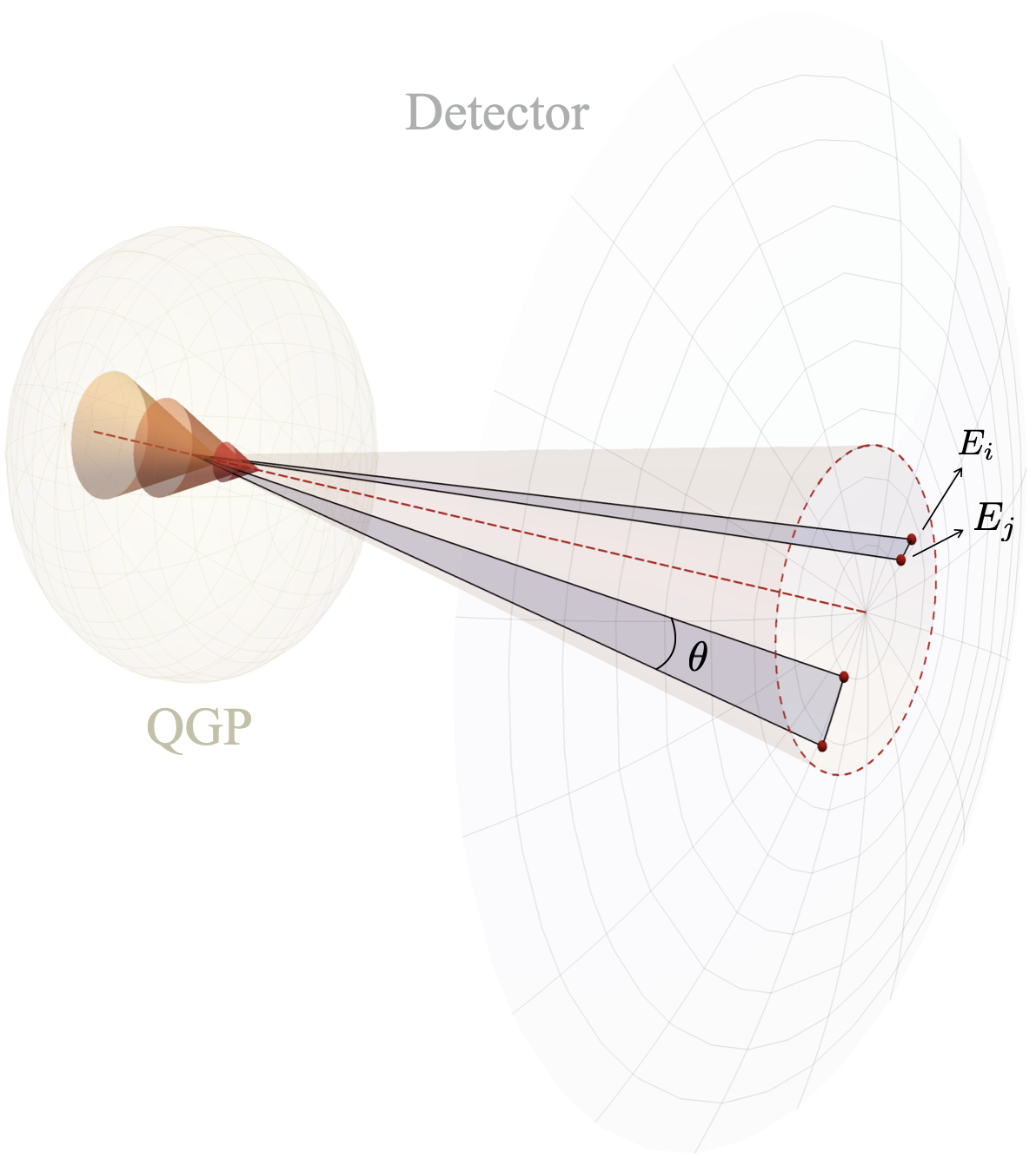}
    \caption{Visualization of the measurement of the 
    medium response sourced by a jet using EECs, see Eq.~\eqref{eq:EEC_exp_definition}. Here the series of colored cones illustrates the medium response, the solid sphere stands for the QGP medium through which the probe propagates, and the hydrodynamical response from the QGP is measured in an idealized, far-away spherical detector.}
    \label{fig:cartoon}
\end{figure}

\noindent\textit{Introduction:} As highly energetic particles, i.e. jets, blaze through the hot quark gluon plasma (QGP) produced in heavy-ion collisions (HIC), they are trailed by an excess of soft particles originating from the background matter~\cite{Casalderrey-Solana:2004fdk,Satarov:2005mv, Chaudhuri:2005vc, Ruppert:2005uz, Casalderrey-Solana:2006lmc}. Since the QGP is expected to be described by relativistic hydrodynamics, some of these particles form a Mach cone speared by the fast parton, see e.g.~\cite{Neufeld:2009ep,Qin:2009uh}, while the remaining form a diffusive wake, see e.g.~\cite{Gubser:2007ga,Chesler:2007sv}, as schematically illustrated in Fig.~\ref{fig:cartoon}. If experimentally observed, this hydrodynamic response would provide yet another evidence for the fluid behavior of the QGP, being hard to mimic with alternative descriptions of the bulk's dynamics, see e.g. \cite{Romatschke:2015dha,Romatschke:2018wgi,Gambini:2016ijt,Hagiwara:2017ofm}. Although considerable effort has been put in characterizing~\cite{Tachibana:2017syd,He:2015pra,Casalderrey-Solana:2020rsj,Yang:2021qtl,Yang:2022nei,Zhou:2024ysb} and measuring~\cite{ATLAS:2020wmg,CMS:2021otx,CMS:2016cvr,CMS:2011iwn,ALICE:2023jye,ATLAS:2024prm, CMS-PAS-HIN-23-006} the medium response to the jet, our theoretical understanding has been mainly driven by numerical studies, while some experimental evidence of the jet wake has been found.

The study of the medium back-reaction has been revamped by the renewed interest in energy correlator (ENC) based jet observables, see e.g.~\cite{Basham:1979gh,Basham:1978bw,Sveshnikov:1995vi,Dixon:2019uzg,Lee:2022ige,Chen:2023zzh,Alipour-fard:2024szj}. These measure the correlations between asymptotic energy fluxes, and have been recently argued to be highly sensitive to the jet induced wake~\cite{Bossi:2024qho,Yang:2023dwc}, seeing a clear imprint of 
the medium back-reaction. Furthermore, ENCs allow for a proper QFT treatment~\cite{Dixon:2019uzg,Chen:2021gdk,Hofman:2008ar}, involving the light-transformed energy momentum tensor (EMT)~\cite{Hofman:2008ar, Sveshnikov:1995vi}:
\begin{align}\label{eq:def_Ec}
 \mathcal{E}(\bn) = \lim_{r\to \infty} r^{2
 } \int_0^\infty dt\, n^i T^{0i}(\bn)\, ,
\end{align}
with $\bn$ corresponding to a point on the celestial sphere, see Fig.~\ref{fig:cartoon}. Assuming that the QGP behaves like a nearly ideal relativistic fluid, the theoretical description of the response in terms of ENCs can be naturally combined with the hydrodynamical evolution of the medium in the presence of an external source:
\begin{align}\label{eq:hydro_eq_1}
    \pt_\m T^{\m\n} = J^{\n}\,.
\end{align}
Here $J$ is a source term, combining the effects of the non-hydrodynamic probes and accounting for the inflow of energy and momentum lost by the jet to the medium. Notice that Eq.~\eqref{eq:hydro_eq_1} is purely classical, and thus it should be understood as only involving the expectation values of the EMT and current operators.

In this paper, we develop an analytic description for the energy and momentum outflow from a hard parton (jet) into the medium, detailing how the QGP back-reacts to this inflow. Our model accommodates recent developments in jet quenching~\cite{Apolinario:2024equ,Cao:2024pxc}, providing an improved description compared to models developed in decades past, see e.g.~\cite{Yan:2017gmv,Iancu:2015uja,Yan:2017rku,Neufeld:2008dx,Neufeld:2009ep,Qin:2009uh}. Aiming to keep the results analytically tractable, we focus on a simple medium model, based on a static homogeneous spherical matter, and allow the hydrodynamic perturbations to reach its boundary, despite that the real fireball matter undergoes hadronization at a finite time. Relying on this setup, we further compute the medium response 
contributions to $\Sigma^{(2)}$, the EEC distribution inside a jet, and show that its leading behavior is highly insensitive to the precise way energy and momentum are transferred from the jet to the medium, and the characteristic features are rather determined by hydrodynamics. \textbf{In general, we establish that the purely medium-response contribution to the EEC parametrically scales as 
$d\Sigma_{\rm response}^{(2)}\propto   d\cos\theta$. This conclusion follows from the classical nature of the hydrodynamic response and the corresponding flux operators.}

\vspace{.2 cm}

\noindent\textit{The jet and its wake:} As a jet propagates through the QGP, it loses energy, via radiative processes, and transfers transverse momentum to the medium, see e.g.~\cite{Casalderrey-Solana:2007knd,Mehtar-Tani:2013pia,Cao:2020wlm} for detailed reviews. These momentum and energy inflow rates are sufficiently small, such that the resulting response of the medium evolution can be considered as a perturbation.
This motivates the linearization of the hydrodynamics set by Eq.~\eqref{eq:hydro_eq_1}, as typically employed in the study of the medium response~\cite{Casalderrey-Solana:2004fdk,Neufeld:2008dx,Neufeld:2009ep,Qin:2009uh,Casalderrey-Solana:2020rsj} in the QGP context. Introducing an EMT perturbation $\delta T^{\m\n}$ around the background, one can write
\begin{align}
\label{hydroEQ}
    \pt_\mu \delta T^{\mu\n} = J^{\n}\,.
\end{align}
These equations describe the evolution of the perturbations in the thermodynamic potentials and flow, and, for a static homogeneous medium, depend on two parameters: the speed of sound $c_s$ and the dissipative damping rate $\Gamma_\eta=\frac{\eta}{w}$, where $\eta$ is the shear viscosity and $w=\epsilon+p=sT$ is the enthalpy. Solving the linearized evolution in Eq.~\eqref{hydroEQ}, for a static background, we readily find the corresponding time-integrated energy flux
\begin{align}
    & \int_0^\infty dt\, \d T^{0 i}(t,\k) = -\frac{i\k^i}{\k^2}\int_0^\infty dt\, J_0 (t,\k)\notag\\
    &\hspace{1.5cm}+\left(\d^{ij}-\frac{k^i k^j}{\k^2}\right)\frac{1}{\Gamma_\eta \k^2}\int_0^\infty dt\,J^j(t,\k)\,,
    \label{T0iimp}
\end{align}
where we have assumed that the source functions quickly fall with time.

Since jets are in general highly collimated objects, with most of the energy and momentum being transported along the jet axis, we describe the jet in terms of a classical current corresponding to a highly energetic (leading) parton propagating along the $z$-axis:
\begin{align}
    & J^{\mu}(\x,t) = \left(\frac{d E}{d t}, \frac{d\p_\perp}{dt}, \frac{dE}{dt}\right)\delta^{(2)}(\x_\perp)\delta(t-z)\, ,
    \label{eq:Jmu1}
\end{align}
where we use bold symbols to represent vectors, and $\perp$ to denote the transverse 2-dimensional subspace. The energy loss rate $dE/dt$ considered here neglects collisional energy loss, since radiative quenching processes dominate at high energy, see e.g. \cite{Wicks:2005gt, Zakharov:2007pj, Qin:2007rn}. In turn, the transverse momentum exchange involves both radiative and elastic contributions, and we will model it as a stochastic quantity with zero mean. Notice that the hydrodynamic equations are in general applicable only at macroscopic distances, and the delta functions in Eq.~\eqref{eq:Jmu1} should be regulated when the response is considered close to the source.

To fully characterize the jet current, one needs to provide explicit prescriptions for the energy and momentum fluxes. For the former, we first note that the energy inflow rate from the energetic parton into the matter differs from the radiative energy loss rate~\cite{Caron-Huot:2010qjx,Jeon:2003gi,Zakharov:1996fv,Baier:1996vi,Arnold:2002ja,Gyulassy:2002yv}, since not all the emitted gluons directly thermalize into the matter. Rather, only gluons whose energy goes down to a scale of the plasma temperature are fully captured by the medium. The branching process controlling the breakdown of the jet energy was first introduced in
~\cite{Baier:2000sb}, corresponding to an inverse Kolmogorov energy cascade, see e.g.~\cite{Berges:2020fwq} for a recent review of thermalization dynamics in QCD. The respective rate of energy inflow into the medium can be obtained from the single gluon inclusive distribution. Its leading behavior can be described by a simplified evolution kernel~\cite{Blaizot:2013hx} and, in the limit of a long and dense medium, it is given by
\begin{align}\label{eq:D}
 D(x,\tau) &= \frac{\tau\,e^{-\frac{\pi \tau^2}{1-x}}}{\sqrt{x}(1-x)^{\frac{3}{2}}}    \Theta(\tau)\Theta(a_\tau L-\tau) + \delta(1-x)\Theta(-\tau)\notag\\
 &\hspace{0cm}+\frac{L\,e^{-\frac{\pi L^2}{1-x}}}{\sqrt{x}(1-x)^{\frac{3}{2}}}   \Theta(\tau-a_\tau L)\, ,
\end{align}
where $x$ is the fraction of the gluon energy $\o$ with respect to the initial jet energy,\footnote{We note that, strictly speaking, this form of the inclusive gluon distribution is only valid in the regime where $\hat q L^2 \gg p_t$. While the high-energy regime can also be taken into account by extending the relevant formalism~\cite{Fister:2014zxa}, to describe medium induced energy loss it is sufficient to consider Eq.~\eqref{eq:D}.} and we have introduced a dimensionless time $\tau \equiv \left(\alpha_s C_A/\pi\right) \sqrt{\hat q /p_t}\,t = a_\tau  t $, with $\hat q$ the jet quenching parameter and $L$ the characteristic size of matter. Notice that this form extends the gluon distribution introduced in~\cite{Blaizot:2013hx,Fister:2014zxa}, by accounting for the finite path of the probe in the medium and allowing for energy to flow to the medium only for $\tau>0$. 
The energy loss rate can be defined from Eq.~\eqref{eq:D} as
\begin{align}
&\frac{1}{p_t}\frac{dE}{d\t}\equiv - \partial_\tau \int^1_{x_0 } dx \,  D(x,\tau) = \frac{2}{\sqrt{\pi}} e^{-(\gamma+\pi)\tau^2} \notag\\
&\hspace{0cm}\times \left( \sqrt{\gamma} + e^{\gamma \tau^2} \pi^{\frac{3}{2}} \, \tau \, {\rm erfc}(\sqrt{\gamma} \tau) \right) \Theta(\tau)\Theta(a_\tau L-\tau) \, , \label{dEdsgen}
\end{align}
where $\gamma = \frac{\pi x_0}{1-x_0}$, and $x_0 p_t\sim T$ is the thermalization bound, reflecting the fact that the gluons with lower energy are considered to be a part of the medium. Notice that for a parton forming a jet, $T\ll p_t$, and one expects that $\gamma \ll \pi$.

The transverse current can be obtained by noticing that as the jet evolves, its constituents scatter multiple times on the medium. At leading order, such interactions are elastic and thus the momentum flowing into the jet must be compensated by an opposite flux which distributes transverse momentum into the matter constituents. Within the same model, the rate of the transverse momentum deposition $\frac{d\p_\perp}{dt}$ can then be described as follows. The path of the leading parton should be divided into steps of a small size, controlled by the hydrodynamic scale, $\D s \sim 1/T$. The momentum deposited into the matter on the $i$th step, $\D \p_\perp^{(i)}$, is a stochastic quantity. As an example, one could take this to be normally distributed over multiple events, $\langle \D p_\perp^{(i)a}\D p_\perp^{(j)b}\rangle_{\rm stoch}=\d^{ij}\d^{ab}\,\Delta s\, \hat{q}/2$, following the form typically considered in the literature, see e.g.~\cite{Barata:2020rdn}. The smoothed momentum inflow rate can be obtained by comparing $\D \p_\perp^{(i)}$ with $\D s$, since the hydrodynamic matter is not expected to resolve short distance features of the process, assuming that $\Delta s \gg \lambda_{\text{mfp}}$ or equivalently that the medium stays densely populated at the scale $\Delta s$ . Other forms of this distribution can be straightforwardly implemented. 

\begin{figure*}[t!]
    \centering
    \includegraphics[width=1.0\textwidth]{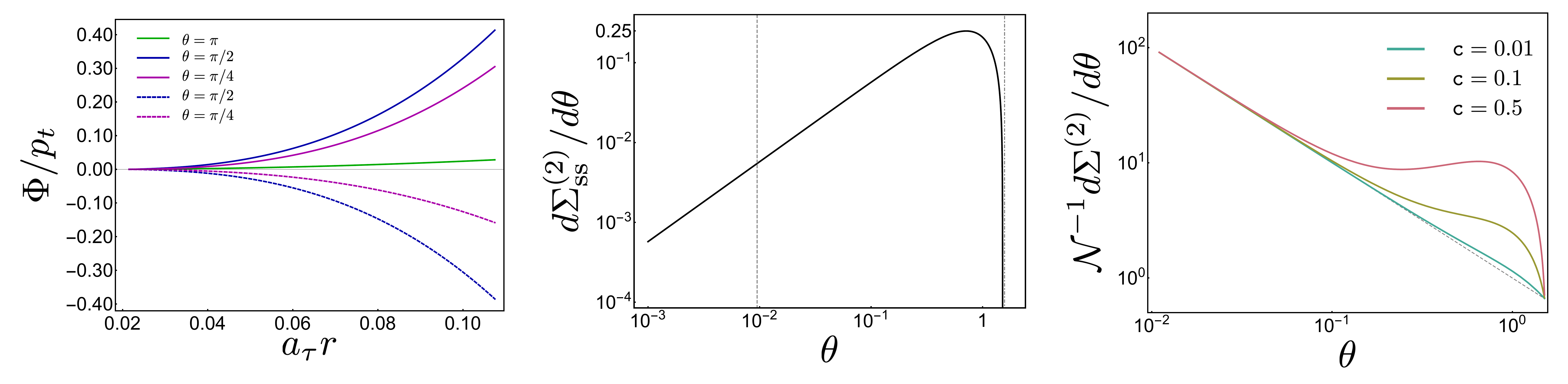}
    \caption{\textbf{Left:} Energy flow $\Phi$ through symmetric spherical surfaces of angular sizes $\theta=\pi,\,\pi/2$, and $\pi/4$ centered around the jet axis (solid lines), as well as through the corresponding symmetric spherical surfaces, centered around the jet axis in the negative direction (dashed lines). \textbf{Center:} Pure
    medium response (soft-soft) contribution to $\Sigma^{(2)}$ with no additional angular cuts imposed, the vertical dashed lines highlight the plotting range of the right panel.
    \textbf{Right:} Full $\Sigma^{(2)}$ distribution, including the soft and hard contributions, within the qualitative description detailed in the text. The different curves correspond to $\mathtt{c}=0.01$ (cyan), $0.1$ (green) and $0.5$ (magenta), see Eq.~\eqref{eq:Sigma2_full}.}
    \label{fig:flux}
\end{figure*}

With this construction, we can now study the features of the medium response within the given event. To gain a basic insight, it is instructive to consider first the distribution of the energy flow $\Phi$ through a distant spherical surface, being trivially connected to the one-point energy correlator. Using the explicit solution in Eq.~\eqref{T0iimp}, we readily find
\begin{align}\label{Eflow}
&\Phi =r^2\int_t\int_{\bn}\,n^i\, T^{0i} \nn 
&= \left(r\cos\psi\,\frac{1-\cos^2\theta}{4\Gamma_\eta }+\frac{1-\cos\theta}{2}\right)\int_s\frac{dE}{ds} + \mathcal{O}\left(\frac{1}{r}\right) \, ,
\end{align}
where $\theta$ controls the size of the symmetric sphere segment, $\psi$ defines the direction to its center, and we have assumed that the energy inflow rate falls quickly with the path. The dominant term in the energy flux at large distances corresponds to the diffusion wake, while the second term originates from the sound modes. Notice that the transverse momentum exchange rate averages to zero in the considered one-point function. 

It is however expected that the leading parton loses only a small fraction of its energy, and the finite size effects cannot be neglected. Indeed, taking $p_t=200\,\text{GeV}$, $x_0=0.001=\frac{T}{p_t}$, $g=2$, $\hat{q}=0.2\,\text{GeV}^2/\text{fm}$, and\footnote{Here, we use the strongly coupled limit for $\eta/s=1/(4\pi)$ appearing in holographic considerations \cite{Policastro:2001yc,Kovtun:2004de}.} $\Gamma_\eta=\frac{1}{4\pi T}$ for illustrative purposes, we find that the characteristic distance $r\sim1/a_\t\simeq 47\,\text{fm}$ of our model is rather large comparing to the typical size of the QGP droplets $r\sim 5\,\text{fm}$. Thus, finite size corrections must be taken into account. The general behavior of the energy flux sourced by the medium response is illustrated in Fig.~\ref{fig:flux} (\textbf{Left}), where we plot the flux through a symmetric spherical segment centered at the jet axis (in positive or negative direction) as a function of the medium radius, for angular sizes $\theta=\pi$, $\pi/2$, and $\pi/4$. Notice that we regulate the short distance behavior of the source by switching off the energy loss rate $1/T$ away from the medium boundary. To obtain the curves in Fig.~\ref{fig:flux}, we numerically solve for $\mathcal{E}(\bn)$ using Eq.~\eqref{T0iimp}, while Eq.~\eqref{Eflow} solely illustrates the leading order behavior of $\Phi$ in $r$. We can readily see that the momentum flux quickly grows in its absolute value in the forward and backward regions, while the full flux through the spherical matter changes much slower. The former effect corresponds to the diffusive wake, pulling the matter along the jet axis, whereas the overall outgoing flux is carried by the sound modes. 

The matter produced in HIC has a finite extension, captured by the medium size in our model, and the hydrodynamic modes cannot propagate beyond it. In what follows, we will assume that the energy flux through the actual detector, placed at much larger distance from the center, is just a linear projection of the energy flux through the medium boundary, assuming that the hadronization is sufficiently isotropic at larger angles.

\vspace{.2 cm}

\noindent\textit{Medium response contribution to $\Sigma^{(2)}$:} Having discussed the basic features and properties of the
medium response, we now consider its contribution to the EEC. In the current construction, where the medium response can be clearly separated from the jet initiating it, it is convenient to divide the EMT into hard (h) and soft (s) parts, classifying its correlations accordingly. Here, we are interested in the medium response, and mainly focus on the soft-soft (ss) contribution to the EEC, only briefly commenting on the hard-soft counterpart later on. In turn, the hard-hard counterpart of the EEC is unaffected by the medium-response, and a more detailed discussion can be found in the literature, see e.g. \cite{Singh:2024vwb,Barata:2023bhh,Yang:2023dwc,Andres:2023xwr,Xing:2024yrb,Devereaux:2023vjz}. Moreover, the medium response contribution to the EMT can be further divided into steady (energy inflow) and stochastic (transverse momentum transfer) parts, and their cross terms with odd powers of $\frac{d\p_\perp}{dt}$ should naturally vanish averaging to zero.

The corresponding $\Sigma^{(2)}$ distribution is commonly defined as
\begin{align}\label{eq:QFT_S2}
     &\frac{d\Sigma^{(2)}}{d\cos \theta} =- \int_{\bn_{1},\bn_{2}}\frac{\langle \mathcal{E}(\bn_1) \mathcal{E}(\bn_2) \rangle}{p_t^2}\delta(\bn_1\cdot\bn_2-\cos\theta)\nn 
    &= - \hat{
     \sum\limits_{\rm events}}\int_{\bn_{1},\bn_{2}}\frac{ \mathcal{E}_{\rm s}(\bn_1) \mathcal{E}_{\rm s}(\bn_2) }{p_t^2}\delta(\bn_1\cdot\bn_2-\cos\theta)\, ,
\end{align}
where in the second line we only keep track of the medium response contribution. Here, we also assume that $\theta\in(0,\pi)$, $\hat{
     \sum\limits_{\rm events}}$ denotes the average over events, and $\mathcal{E}_s(\bn)$ is the energy flux along the $\bn$ direction due to the soft particles in a given event. It should be stressed that the hydrodynamic modes are classical, and, at the simplest level, focusing on the steady contributions, one has to consider the classical limit of the corresponding EEC. It is also useful to illustrate the explicit form of this object for an ensemble of particles:
\begin{align}\label{eq:EEC_exp_definition}
     \frac{d\Sigma^{(2)}}{d\cos \theta} =- \hat{
     \sum\limits_{\rm events}} \sum_{i\neq j}\frac{E_iE_j}{p_t^2}\d(\bn_i\cdot\bn_j-\cos\theta)
     \, ,
\end{align}
which can be directly related with the EEC distribution commonly used on the experimental side $d\Sigma^{(2)}/d\theta =\hat \sum_{\rm events} \sum_{i\neq j}\frac{E_iE_j}{p_t^2}\d(\theta_{ij}-\theta)$~\cite{Tamis:2023guc,CMS:2024ovv}, with $\theta_{ij}$ the relative angle between the particle pair $ij$. Notice that the information about mutual correlations is now hidden in the particular distribution within the ensemble of particles, and one has to further specify whether the particles are chosen over the whole ensemble or only its particular part (within the whole event or within jets).

The energy flux due to the medium response sourced by a jet propagating along the $z$-axis is azimuthally symmetric. Thus, rotating the coordinate system to define the angles of one $\bn$ around the other, 
we readily find
\begin{align}\label{eq:EEC_final}
&\frac{d\Sigma^{(2)}_{\text{ss, steady}}}{d\cos \theta} =- \hat{
     \sum\limits_{\rm events}} \frac{2\pi}{p_t^2}\int_{\Theta,\phi}\mathcal{E}_{\text{s}}(\cos\Theta)\sin\Theta\notag\\
&\hspace{2cm}\times\mathcal{E}_{\text{s}}(\sin\Theta\sin\theta\cos\phi+\cos\Theta\cos\theta)\, ,
\end{align}
where $\phi$ is the azimuthal angle of, for example, $\bn_1$ around $\bn_2$ and $\Theta$ is the polar angle of $\bn_2$ defined with respect to the jet axis. We also assume here that any additional constraints on the integration limits, e.g. the ones defining the jet cone, are incorporated into the energy fluxes. Further expanding this for small angles, one may notice that $\frac{d\Sigma^{(2)}_{\text{ss, steady}}}{d\theta} = c_1 \theta+\mathcal{O}(\theta^3)$ with\footnote{Notice also that this scaling law is also observed in the \textit{hadron gas} region of the EEC in the vacuum, see e.g.~\cite{Komiske:2022enw}.}
\begin{eqnarray}\label{eq:c1_new}
c_1=\hat{
     \sum\limits_{\rm events}}\frac{(2\pi)^2}{p_t^2}\int_{\Theta}\mathcal{E}_{\text{s}}^2(\Theta)\sin\Theta\sim\frac{r^2}{\Gamma_\eta^2}\frac{\hat{q} r^2}{p_t}\,,
\end{eqnarray}
where the dependence on $\hat q$ is connected to the energy loss rate in Eq.~\eqref{dEdsgen}. This angular scaling is a general feature of the EEC of classical fluxes, which are uncorrelated otherwise, e.g. as long as the angles are large enough to ignore the hadronization effects. In fact, within the simple model considered here, we cannot account for hadronization consistently, but, on general grounds, the spectrum is expected to quickly fall for small enough angles, and in what follows we will ignore the region where $\theta \ll {\Lambda_{\rm QCD}}/T$. 

Turning to the stochastic counterpart of the medium response, it is instructive to consider the small angle behavior of the stochastic contribution to the EEC. For the (ultra)local form of the stochastic correlations of the transverse currents the distribution is regular, and one readily finds
\begin{align}c_{1\perp}=\hat{
     \sum\limits_{\rm events}}\frac{(2\pi)^2}{p_t^2}\int_{\Theta}\left\langle\mathcal{E}_{\text{s}}^2(\Theta)\right\rangle_{\rm stoch}\sin\Theta\sim \frac{r^2}{\Gamma_\eta^2}\frac{\hat{q}r}{p_t^2}\, .
\end{align}
Thus, we conclude that the only non-trivially correlated contribution to the EEC in our model exhibits the same small angle behavior as the classical counterpart, $\frac{d\Sigma^{(2)}_{\text{ss}}}{d\theta} \simeq (c_1+c_{1\perp}) \theta$ for small angles. It is however suppressed by $p_t r$, and the particular relative numerical coefficient depends on the explicit form of the energy loss rate.\footnote{We note that the scalings for the steady and stochastic parts of the numerical prefactor are well expected. The steady piece is determined by the form of Eq.~\eqref{eq:D}, where the characteristic energy scale, $\hat q r^2$, is determined by the LPM effect~\cite{Landau:1953gr}. The stochastic term, due to its Gaussian form, has a diffusive nature and thus it must scale with the characteristic average squared momentum, $\hat q r$~\cite{Barata:2020rdn}.}

Focusing again on a finite size system, we project the flux at its boundary to the actual detector, and switch off the energy loss processes $1/T$ away from the boundary, accounting in this way for the short distance cutoff in Eq.~\eqref{eq:Jmu1}. Taking the same parameter values as in the results shown in Fig,~\ref{fig:flux} (\textbf{Left}) we obtain the full soft-soft contribution to the EEC, see Fig.~\ref{fig:flux} (\textbf{Center}). Here, we have included the effects due to both the steady part of the source $dE/ds$, leading only to a classical contribution in Eq.~\eqref{eq:EEC_final}, and due to the stochastic counterpart of the source $d\p_\perp/ds$, resulting in a non-trivially correlated contribution to the EEC. We have found that the latter is significantly suppressed for our parametric choice, as expected from the general discussion above.\footnote{We iterate that to obtain these curves we do not rely on the small angle expansion shown above.}

Thus, we conclude that the characteristic behavior of the soft-soft contribution to the EEC only weakly depends on the particular form of the energy flux distribution on the sphere. Therefore, one may expect that the mixed hard-soft contribution to the EEC will be similar to the soft-soft contribution, as long as the soft part of the stress energy tensor is considered at the classical level,\footnote{In general, the medium response depends on the particular form of the hard source at the operator level, i.e. $T^{0i}_{\rm s}(T^{\a\b}_{\rm h})$. One might attempt at building a description based on separation of scales, and compute such \textit{quantum} corrections in perturbative QCD. We leave such an exercise for future work.} i.e. $\langle \mathcal{E}_{\rm h} \mathcal{E}_{\rm s} \rangle=\hat
     \sum_{\rm events} \mathcal{E}_{\rm h}  \mathcal{E}_{\rm s} $, and at least for sufficiently small angles. Notice that the corresponding stochastic contribution is also expected to be suppressed as in the soft-soft case, since the transverse momentum exchange between the hard and soft components is balanced, closely resembling the soft-soft contribution. This observation allows to qualitatively capture the collinear limit of $\Sigma^{(2)}$ distribution, excluding the non-perturbative small angle regime, while including all possible hard and soft 
terms\footnote{Here we neglect the perturbative modifications associated to the jet-medium interaction, and we only keep the leading order (rescaled) vacuum result.} by writing 
\begin{align}\label{eq:Sigma2_full}
\frac{d\Sigma^{(2)}}{d\theta} = \mathcal{N}\left[\frac{1}{\theta} + \mathtt{c} \left(\frac{2p_t}{\Delta p_t}\right)\frac{d\Sigma^{(2)}_{\rm ss}}{d\theta}\right] \,.
\end{align}
where $\mathcal{N}$ and $\mathtt{c}$ are numerical parameters and the leading term emulates the perturbative QCD vacuum result~\cite{Dixon:2019uzg}. The soft-soft part of the EEC scales as $\Delta p_t^2/p_t^2$, where $\Delta p_t$ is the energy lost by the jet, and one may expect the hard-soft correlations to be enhanced by a factor of $2p_t/\Delta p_t$, introduced explicitly. Since the overall normalization of the curves is not under theoretical control within our model, we focus only on the shape of the characteristic distribution, normalizing in what follows by $\mathcal{N}$.

In Fig.~\ref{fig:flux} (\textbf{Right}), we show the full EEC distribution obtained in the our model, using three values of $\mathtt{c}=0.01$, $0.1$, and $0.5$. As expected, the larger medium response contribution leads to a characteristic enhancement of the EEC at larger angles. We note in passing that this feature, at a qualitative level, closely resembles the medium induced enhancement obtained solely from perturbative QCD considerations, as also seen in~\cite{Yang:2023dwc}, and further theoretical developments are needed to separate the medium response effects from those associated to the perturbative jet cascade in the HIC context. However, if we were to interpret that behavior in \cite{Yang:2023dwc} within our hydrodynamic consideration, we would gather that this feature is sourced by hard-soft correlations.

 \begin{figure}[t!]
     \centering
     \hspace{-0.5cm}\includegraphics[width=0.5\textwidth]{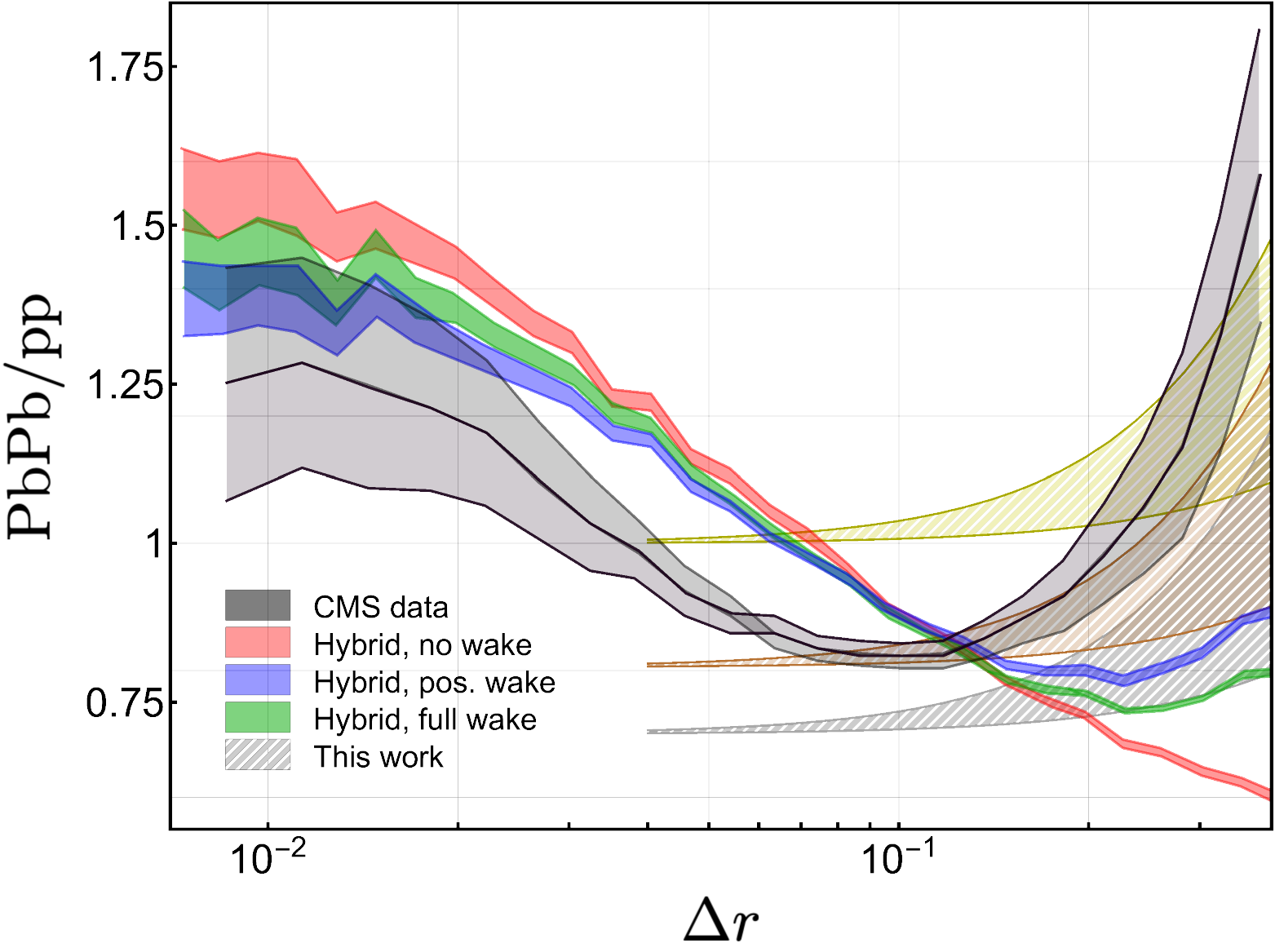}
     \vspace{-0.25cm}
     \caption{Comparison between the results from this work and data from the CMS experiment (central gray bands) -- see Fig.~7 (top left) in~\cite{CMS:2024ovv} -- for the PbPb to pp ratio of $d\Sigma^{(2)}/d\Delta r$, where $\Delta r$ is the angular separation scale. The bands obtained from the Hybrid model~\cite{Casalderrey-Solana:2014bpa} are also shown: red (no wake), blue (positive wake), and green (full wake). Since we do not control the behavior at small angles, where hadronization effects become important, we show three parametric choices in Eq.~\eqref{eq:Sigma2_full} with bands set by $0.1<\mathtt{c}<0.3$: one where the baseline assumes no difference between PbPb and pp at small angles (gold), one which captures the intermediate angular behavior of data (copper), and one which best matches the Hybrid model at large angles (silver). The Hybrid model curves are obtained from~\cite{CMS:2024ovv}.}
     \label{fig:data_comp}
 \end{figure}

To conclude the discussion on the two point correlator, we compare our theoretical results to the publicly available data~\cite{CMS:2024ovv} for the PbPb to pp ratio of the EEC. In particular, in Fig.~\ref{fig:data_comp} we show the data presented in Fig.~7 (top left) of~\cite{CMS:2024ovv}, along with the theoretical predictions obtained from the Hybrid model~\cite{Casalderrey-Solana:2014bpa,Casalderrey-Solana:2015vaa,Casalderrey-Solana:2016jvj} presented in the same work, for mid-rapidity jets with energies in the bin $120 \, {\rm GeV}<p_t< 140 \, {\rm GeV}$, with a jet radius of $R=0.4$, and for the most central collisions, where our model can be suitably applied.\footnote{We note that within our model we can accommodate different $p_t$ bins and experimental settings by taking different ranges for $\mathtt{c}$ and adjusting $r$.} Since we cannot describe the small angle behavior of the ratio, being outside of the validity range of the model, we show three theory bands which differ on the choice of the \textit{starting point}: the gold band corresponds to the choice where at small angles the PbPb and pp EEC distributions exactly match (not favored by data), the copper band corresponds to the choice of taking the value of the ratio to better match the data band, and the silver band is set to better match the Hybrid model curves at larger angles. The overall qualitative features seen in the data are compatible with the hydrodynamic results obtained within our treatment, as well as with the Hybrid model bands. Since in our model the contribution from the jet evolution is not fully taken into account, i.e. no medium modifications to the jet substructure, one may conclude that the behavior seen in the data is compatible with an interpretation where the large angle enhancement is driven solely by hydrodynamics and its (classical) interplay with the hard component, rather than by perturbative branchings. A proper separation of these two contributions requires a more in-depth study, which we leave for future work. 

\vspace{.2 cm}

\noindent\textit{Conclusions and discussion:} In this work, we have solved the linearized hydrodynamics of the QGP bulk in the presence of a localized excitation representing the passage of a jet. Using a simple model for the perturbation, based on jet quenching theory, we have computed the pure medium response contributions to the energy flux and EEC distribution as well as discussed the characteristic form of the cross contributions due to the mixing of the hydrodynamic response and leading jet energy fluxes.

The main conclusion from this study is illustrated in Fig.~\ref{fig:flux} (\textbf{Right}): the jet induced response leads to an enhancement of the EEC distribution at large angles, with a characteristic behavior which is universal when ignoring quantum correlations. Indeed, the small angle behavior of the soft-soft contribution to the EEC in Fig.~\ref{fig:flux} (\textbf{Center}) is determined solely by the geometric correlations in Eq. \eqref{eq:EEC_final}, for further discussion see also \cite{Barata:2025fzd}. In turn, more quantitative features of this contribution cannot be reliably extracted without larger-scale simulations, and we constrain ourselves to a specific set of parameters used for illustrative purposes. A more detailed investigation of its dependence on the full parameter phase space, as well as an extension to account for realistic medium evolution, is left for future work. 

From a phenomenological point of view, in Fig.~\ref{fig:data_comp} we argue that the response's imprint can be visible in this observable, directly competing with the perturbative contributions. These observations are in qualitative agreement with the full scale numerical study of the response in~\cite{Yang:2023dwc}. Even though we have only studied the more suppressed soft-soft contributions, we argue, on general grounds, that our conclusions also apply to the dominant medium response dependent imprint in the EEC, associated with hard-soft pairings.

The present approach can be naturally extended to provide a better modeling of jets and their wakes. One important aspect not taken into account here is the effects of jet substructure fluctuations on the medium response.
Still, in a classical setting, one way to describe this aspect is to consider a two-prong energy-momentum inflow current $  J_{(2)}^\mu(\psi) = J^\mu(t,\psi) +   J^\mu(t,-\psi)$, with $\psi$ being an opening angle. If we rotate the source around an axis in the transverse plane for a small angle, we find
\begin{align}\label{eq:2JJ}
J_{(2)}^\mu(t,\psi) &= \left(\frac{d E}{d t}, \frac{d\p_\perp}{dt}+\bn \frac{dE}{dt} \psi, \frac{dE}{dt}-\left(\frac{d\p_\perp}{dt}\cdot\bn\right)\psi \right)\nn 
&\times \delta^{(2)}(\x_\perp- \bn\, \psi\, t)\delta(z-t) \, ,
\end{align}
where $\bn$ is a unit vector perpendicular to the rotation axis. Repeating the above considerations, weighting this contribution by the perturbative probability to produce the corresponding two-prong state, one may access the mutual correlations between the modes sourced by
the two prongs. To go beyond this \textit{semi-classical} set up, one can use the wide separation of scales between hydrodynamic modes and the jet constituents to treat the jet source current in perturbative QCD, as detailed above, see also~\cite{Pablos:2024muu}. Another interesting and straightforward application of the present model is the computation of more differential~\cite{Barata:2023zqg,Kang:2023big} and higher point correlators~\cite{Chen:2019bpb,Chen:2022swd}, where the imprint of the response's shape could become more transparent~\cite{Bossi:2024qho}. More, it would be valuable to further understand the interplay between our model of the medium's response and the structure of matter resolved by the jet, see e.g. \cite{Sadofyev:2021ohn,Barata:2024xwy,Boguslavski:2023alu, Ipp:2020nfu, Avramescu:2023qvv, Carrington:2022bnv,Avramescu:2024xts,Kuzmin:2023hko}.

Although the presented analytical approach cannot reach the degree of sophistication of full scale numerical simulations of jets and their wakes~\cite{Casalderrey-Solana:2020rsj,Yang:2021qtl}, we have identified several universal features of how the medium response imprints on the jet EEC, and clarified the corresponding theoretical and phenomenological aspects.\footnote{Note that other universal aspects of the jet EEC have also been recently explored, see e.g.~\cite{Liu:2024lxy,Barata:2024wsu}.} A clear understanding of these aspects is essential to achieve a proper and complete theoretical description of this observable, which can only then be compared to data.

\vspace{.5 cm}

\noindent\textit{Acknowledgments:} We thank Paul Caucal, Jorge Casalderrey-Solana, Daniel Pablos, Yeonju Go, Weiyao Ke, Krishna Rajagopal, Alba Soto-Ontoso, Robert Szafron, and Xin-Nian Wang for multiple discussions. We thank  Jorge Casalderrey-Solana, Daniel Pablos, and Krishna Rajagopal for helpful comments on the manuscript. The work of AVS is supported by Fundacao para a Ciência e a Tecnologia (FCT) under contract 2022.06565.CEECIND. AVS and JGM were partly supported by the European Research Council project ERC-2018-ADG-835105 YoctoLHC. The work of MVK is partially funded by the grant of the Foundation for the Advancement
of Theoretical Physics “BASIS”.  

\newpage

\bibliographystyle{bibstyle}
\bibliography{references}

\end{document}